\documentclass[12pt,showpacs,preprintnumbers,aps,prl,groupedaddress]{revtex4-1}

\usepackage{amssymb}
\usepackage{amsmath}
\usepackage{subfigure}
\usepackage{graphicx}
\usepackage{color}
\begin{document}
\preprint{}

\title{Active turbulence in active nematics\\
\small{Lecture notes from the The Geilo School 2015}}
\author{Sumesh P. Thampi}
\affiliation{Department of Chemical Engineering, Indian Institute of Technology Madras, Chennai 600036, India}

\author{Julia M. Yeomans}
\affiliation{ The Rudolf Peierls Centre for Theoretical Physics, 1 Keble Road, Oxford, OX1 3NP, UK}
\email[]{j.yeomans1@physics.ox.ac.uk}
\homepage[]{http://www-thphys.physics.ox.ac.uk/people/JuliaYeomans/}

%

%

%
\begin{abstract}{
Dense, active systems show active turbulence, a state characterised by flow fields that are chaotic, with continually changing velocity jets and swirls. Here we review our current understanding of active turbulence. The development is primarily based on the theory and simulations of active liquid crystals, but with accompanying summaries of related literature.
}
\end{abstract}
%
\maketitle
\section{Introduction}
\label{intro}

Active systems are states of matter that operate out of equilibrium. Examples, which span a wide range of length scales, include cytoskeletal elements such as actin and microtubule networks powered by molecular motors, cells and bacteria,  flocks of birds and schools of fish \cite{Marchetti2013, Sriram2010, Ganesh2011}. In all these cases, energy is input locally, at the level of the individual constituents, thus driving the system into a dynamic state. The local driving differentiates active matter from more traditional examples of non-equilibrium systems where a driving force is applied externally, on the entire system or at the boundaries. Active systems are not exclusively of natural or biological origin; they can be realised synthetically. Examples include granular rods vibrated on a solid surface, Janus catalytic particles whose asymmetry causes self propulsion and light responsive colloids \cite{Narayan2007,Ebbens2015,Palacci2014}.

\begin{figure}
\includegraphics[trim = 0 175 0 0, clip, width = 0.75\columnwidth]{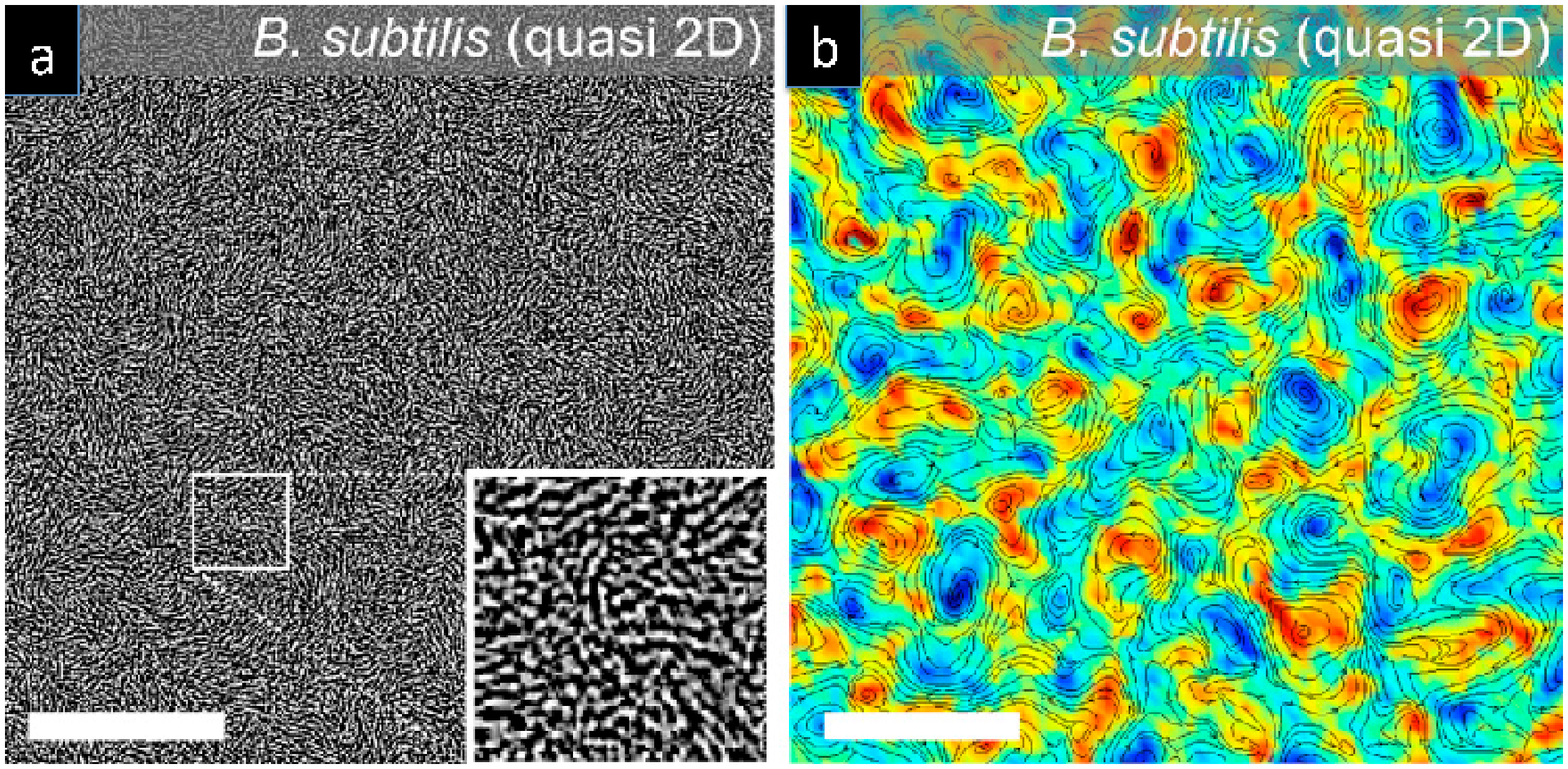}
\includegraphics[trim = 0 160 0 0, clip, width = 0.75\columnwidth]{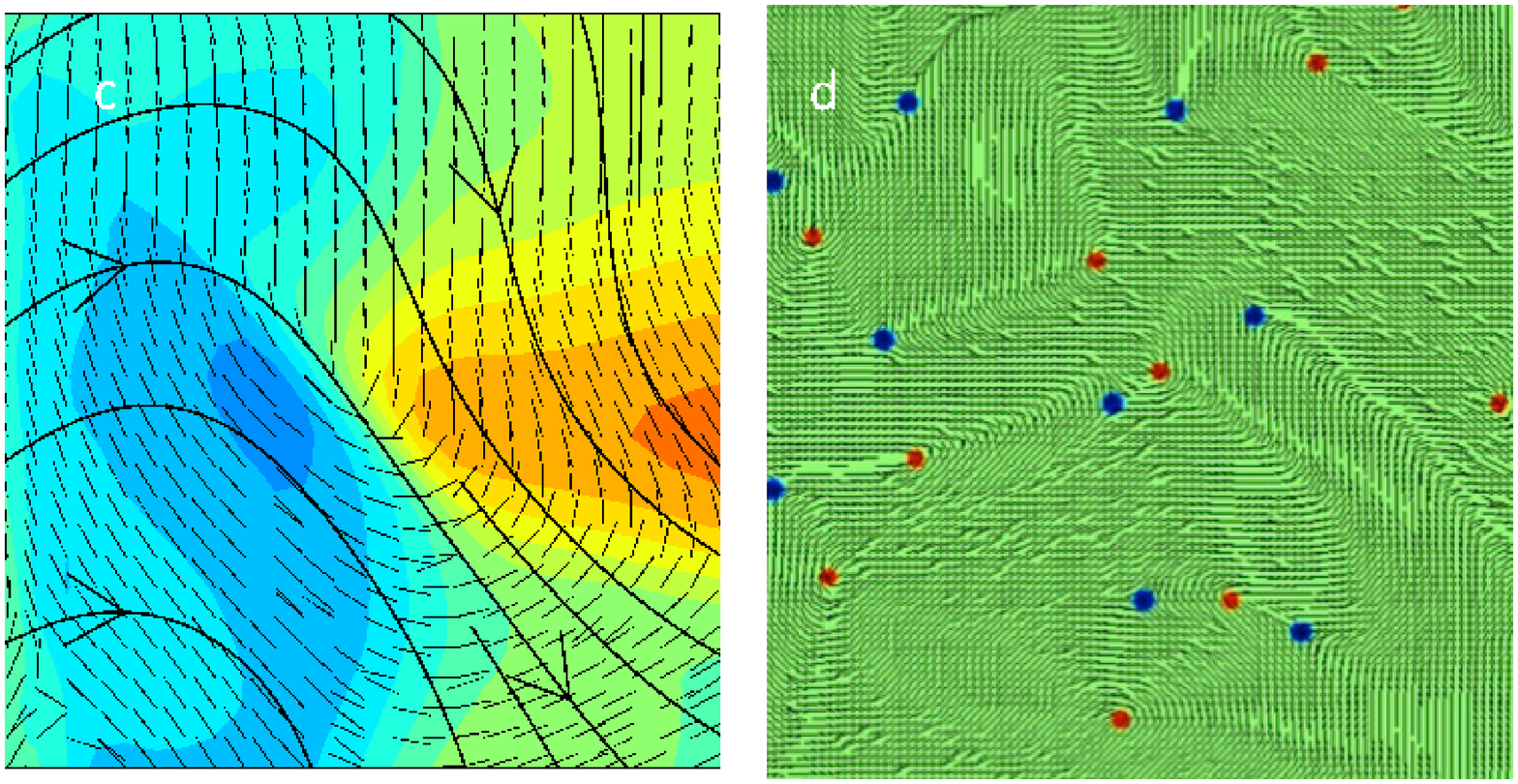}
\includegraphics[trim = 180 215 190 20, clip, width = 0.475\columnwidth]{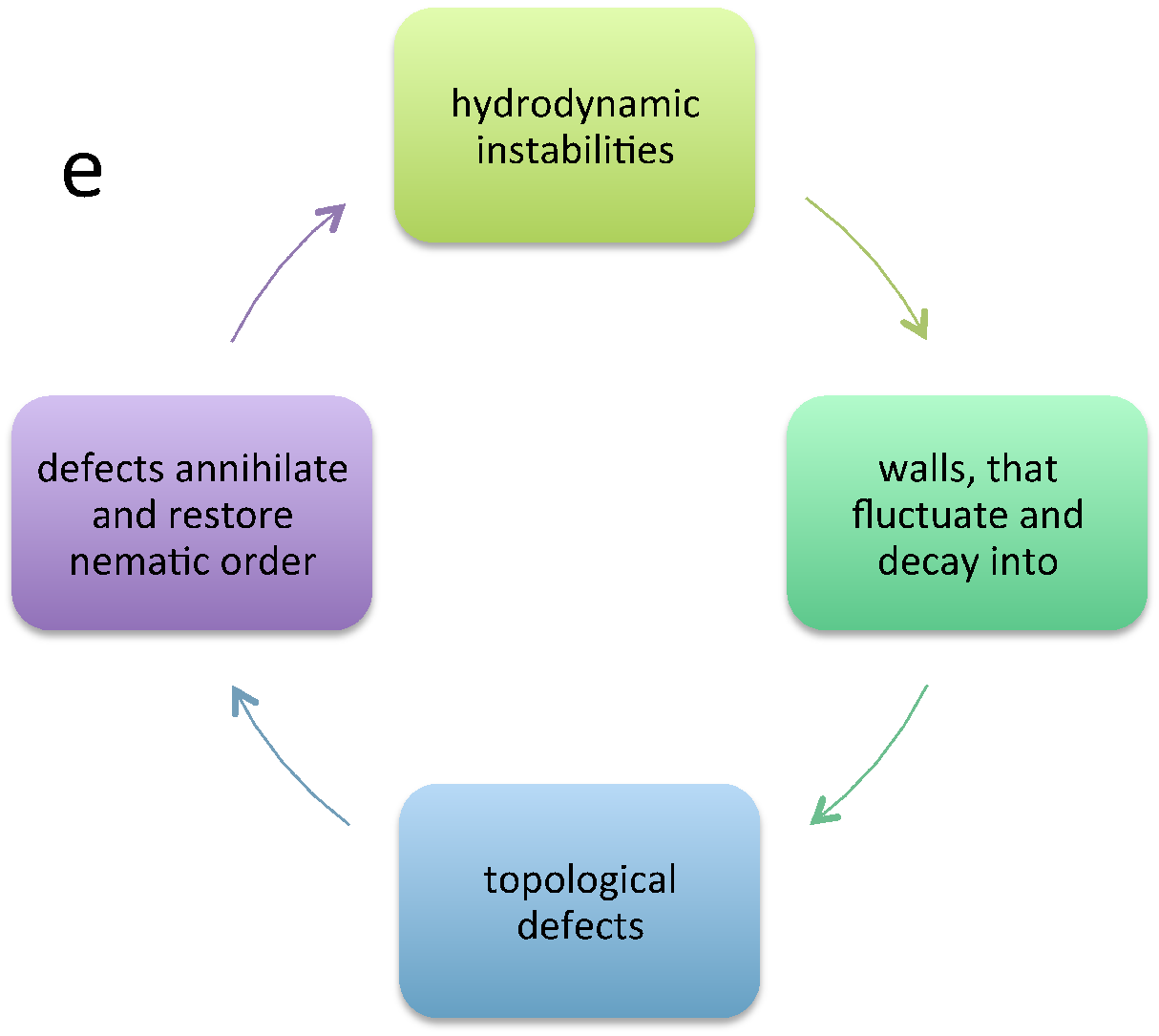}
\label{flowfig}
\caption{Active turbulence. Experiments: (a) Dense suspension of swimming bacteria. (b) The associated vorticity field. Red (blue) regions correspond to high positive (negative) vorticity. After \cite{Julia2012}. Simulations: (c) Active turbulence at a smaller length scale showing a wall. The bend deformation drives a velocity jet (black flow lines, with arrows). (d) $+1/2$ (red) and $-1/2$ (blue) topological defects. (e) Summary of the mechanisms underlying active turbulence. }
\end{figure}

Active particles convert energy into mechanical motion and, as the particles move, they can interact mechanically and  hydrodynamically. As the number density of active particles increases, the interactions become stronger and may dominate over the individuals' dynamics. This can result in states such as active, or mesoscale, turbulence, shown in Fig.~\ref{flowfig}(a) for a dense suspension of bacteria. The velocity field set up by the swimming bacteria has regions of high vorticity, with vortices on a length scale $\sim$ ten times that of an individual bacterium (Fig.~\ref{flowfig}(b)). It is highly chaotic, with flow vortices, swirls and velocity jets continually being created, and destroyed.

Active turbulence has been observed in several experiments.  Wu and Libchaber \cite{Libchaber2000} reported that in two dimensional films of \textit{Escherichia coli}, active turbulence caused tracer particles to behave as if they were in a Brownian bath. The tracers showed a crossover from super diffusive to diffusive behaviour with increasing time.  Later, several studies \cite{Kessler2004, Sokolov2007, Zhang2009, Liu2012, Julia2012, Aranson2012, Jorn2013} measured the velocity statistics of active turbulence produced by the collective motion of \textit{Bacillus subtilis}.  Velocity fields measured in a monolayer of endothelial cells residing on a substrate showed long range hydrodynamic flows, again characterised by localised vorticity \cite{Poujade2007, Petitjean2010, rossen2013}. 
Experiments measuring the velocity-velocity correlation function in suspensions of microtubule bundles and kinesin molecular motors showed that the decay length  of the correlations was insensitive to the ATP concentration \cite{Dogic2012}. Moreover these experiments revealed another important feature of active turbulence: Topological defects are created, transported, and annihilated by the active stresses. Further investigations showed that the topological defects attained nematic order on reducing the system density \cite{Decamp2015}. Confinement of this active liquid on the surface of a vesicle showcased the dynamics of defects constrained on a sphere \cite{Keber2014}.  All these examples can be classified as `wet' systems due to the importance of hydrodynamic effects. However,  active-turbulent-like behaviour and topological defects have also been observed in `dry' systems where hydrodynamics does not play an important role, examples include fibroblast cells \cite{duclos2014} and vibrated granular rods \cite{Narayan2007}. 

Broadly speaking, there are two different theoretical approaches to describing active turbulence. The first considers collections of individual active particles or `swimmers'  \cite{Graham2005, Shelley2008, Shelley2008b, Ganesh2011, Ezhilan2013, Shelley2012, Shelley2014}. Each active entity can be approximated as a force dipole and generates a stresslet velocity field. Simulations show how the active particles interact hydrodynamically to mirror the collective behaviour and large scale flow fields seen in bacterial suspensions. This literature also develops a kinetic theory to yield governing equations that describe the probability distributions of the active particles. 

The second approach is active liquid crystal theory where the continuum equations of motion follow from symmetry arguments\cite{Sriram2002, Davide2007, Sriram2010, Giomi2011, Giomi2013, Marchetti2013, ourpta2014,Giomi2014,Putzig2015}. The active flow field has nematic symmetry, so the most widely used form of the hydrodynamic equations are those of nematic liquid crystals supplemented by an active stress. Equations with polar symmetry have also been studied \cite{Marchetti2013} and purely phenomenological models can also produce the patterns observed in active turbulence \cite{Julia2012, Oza2015}.

The aim of this tutorial review is to describe the active nematic approach to active turbulence. We first discuss in detail how the active stress is incorporated into the equations of motion of the passive nematics, and then consider the hydrodynamic instability of the active equations which gives rise to active turbulence.  Numerical solutions of the equations of motion allow us to describe active turbulence in terms of the cycle depicted in Fig 1E: hydrodynamic instability $\Rightarrow$ formation of walls in the director field $\Rightarrow$ topological defects generated at the walls $\Rightarrow$ defects move and annihilate restoring nematic order $\Rightarrow$ which again undergoes the hydrodynamic instability.
We conclude by summarising related research and open questions.

\section{Theory}


We first summarise the equations of motion of a passive nematic liquid crystal, following the approach of Beris and Edwards. The nematohydrodynamic equations can be derived from linear irreversible thermodynamics with phenomenological models that assume relation between fields and currents, or by using a Poisson bracket formalism \cite{Berisbook, DeGennesBook,Sriram2002, Marchetti2013} .

The variable used to characterise the magnitude and orientation of the nematic order is a symmetric, traceless, second rank tensor $\mathbf{Q}$. If $\mathbf{n}$ represents the director field, $\mathbf{Q} = \frac{q}{2} ( 3\mathbf{nn} - \mathbf{I})$ for uniaxial nematics where $q$ is the largest eigenvalue of $\mathbf{Q}$ and $\mathbf{I}$ is the identity tensor. The nematohydrodynamic equations of motion, which  govern the evolution of  $\mathbf{Q}$, together with the two conserved variables, density $\rho$ and momentum density $\rho \mathbf{u}$, are
\begin{align}
(\partial_t + u_k \partial_k) Q_{ij} - S_{ij} &= \Gamma H_{ij},
\label{eqn:lc}\\
\rho (\partial_t + u_k \partial_k) u_i &= \partial_j \Pi_{ij};~~~~\partial_i u_i = 0.
\label{eqn:ns}
\end{align}

Consider first eq.~(\ref{eqn:lc}) for the order parameter. $S_{ij}$ is the generalised nonlinear advection term which describes how the nematic order responds to velocity gradients. It takes the form
\begin{align}
S_{ij} = (\lambda E_{ik} + \Omega_{ik})(Q_{kj} + \delta_{kj}/3) + (Q_{ik} + \delta_{ik}/3)&(\lambda E_{kj} - \Omega_{kj})\nonumber\\ &- 2 \lambda (Q_{ij} + \delta_{ij}/3)(Q_{kl}\partial_k u_l)
\end{align}
where $E_{ij} = (\partial_i u_j + \partial_j u_i)/2$ is the strain rate tensor, characterising the deformational component, and $\Omega_{ij} = (\partial_j u_i - \partial_i u_j)/2$ is the vorticity tensor, characterising the rotational component, of the flow field. The vorticity vector is defined as the curl of the velocity field, $\boldsymbol{\omega} = \nabla \times \mathbf{u}$ and it is related to the vorticity tensor by $\epsilon_{ijk}\omega_k = -2\Omega_{ij}$ where $\boldsymbol{\epsilon}$ is the Levi-Civita symbol. 
The degree of coupling between $\mathbf{Q}$ and the velocity gradients is controlled by the alignment parameter $\lambda$ \cite{DeGennesBook, Scott2009}. 
For a uniaxial liquid crystal with an order parameter of constant magnitude $q$ 
we can define $\tilde{\lambda}=(3q+4)\lambda/9q$ \cite{Davide2007}. Particles with $|\tilde{\lambda}| > 1$ align in a simple shear flow, whereas those with $|\tilde{\lambda}| <1 $ tumble in the flow.

The term on the rhs of eq.~(\ref{eqn:lc}) describes the relaxation of $\mathbf{Q}$ to the minimum of a free energy $\mathcal{F}$. This dynamics is driven by the molecular field, the variational derivative of the free energy,
\begin{align}
H_{ij} = -\frac{\delta \mathcal{F}}{ \delta Q_{ij}} + \frac{\delta_{ij}}{3} {\rm Tr} \frac{\delta \mathcal{F}}{ \delta Q_{kl}},
\end{align}
at a rate determined by the rotational diffusion coefficient $\Gamma$. The free energy is usually taken to be of the Landau-de Gennes form. There is a bulk contribution chosen as a truncated polynomial expansion in products of $\mathbf{Q}$ with scalar symmetry \cite{Berisbook, DeGennesBook}, together with a term in gradients in $\mathbf{Q}$ to model the elastic free energy that arises from distortions in the director field and changes in the magnitude of the order parameter.
\begin{align}
\mathcal{F} = \frac{A}{2} Q_{ij} Q_{ji} +\frac{ B}{3} Q_{ij} Q_{jk} Q_{ki} +\frac{C}{4} (Q_{ij} Q_{ji})^2 + \frac{K}{2} (\partial_k Q_{ij})^2,
\end{align}
where $A, B$ and $C$ are phenomenological coefficients which are functions of concentration and temperature and we have written a single elastic constant ($K$) approximation.

We now turn to eq. (\ref{eqn:ns}) which describes the evolution of the velocity field. Written in this compressed form it closely resembles the Navier-Stokes equations.  The complications that arise from the viscoelasticity of the nematic fluid appear in the stress tensor  $\Pi_{ij}$ . This includes the usual viscous stress,
\begin{equation}
\Pi_{ij}^{viscous} = 2 \mu E_{ij} ,
\label{eqn:viscstress}
\end{equation}
where $\mu$ is the Newtonian viscosity of the suspension. However account must also be taken of 
elastic stresses, often referred to as the `back-flow', which arise because changes in the orientational order generate forces on the fluid:
\begin{align}
\Pi_{ij}^{passive}=&-P\delta_{ij} + 2 \lambda(Q_{ij} + \delta_{ij}/3) (Q_{kl} H_{lk})
-\lambda H_{ik} (Q_{kj} + \delta_{kj}/3)  \nonumber\\ &- \lambda (Q_{ik} + \delta_{ik}/3) H_{kj}
-\partial_i Q_{kl} \frac{\delta \mathcal{F}}{\delta \partial_j Q_{lk}} + Q_{ik}H_{kj} - H_{ik} Q_{kj}
\label{eqn:passstress}
\end{align}
where $P = \rho T - \frac{K}{2} (\partial_k Q_{ij})^2$  is the modified bulk pressure.
The back-flow can affect the director dynamics, for example accelerating defect annihilation \cite{Julia2002} and modifying the motion of active defects \cite{Giomi2013}. It is, however, usually small compared to imposed or active stresses.

These are the stress contributions in a passive nematic. The only new term needed in the hydrodynamic equations of an active nematic is an additional contribution to the stress tensor
\begin{equation}
\Pi_{ij}^{active} = -\zeta Q_{ij} ,
\label{eqn:actistress}
\end{equation}
where $\zeta$ is a measure of the strength of the activity. The active stress, introduced in \cite{Sriram2002}, occurs because to leading order active particles act as force dipoles. We will derive this term in the next section.

More details about the equations of motion and their application to passive and active systems can be found in  \cite{Berisbook, DeGennesBook, Denniston2001, Denniston2004, Davide2007, Cates2008, Orlandini2008, Henrich2010, Suzanne2011, Miha2013}.

\subsection{Constitutive relation of a suspension of active particles}

Consider a suspension of identical active particles, and let $\mathbf{U}, \mathcal{P}$ and $\boldsymbol{\sigma}$, respectively, represents the velocity, pressure and stress field in the active medium. In a volume, $V$, containing a sufficient number of active particles to perform averaging, the mean value of the stress tensor may be written \cite{BatchelorBook, Batchelor1970}
\begin{align}
\bar{\sigma}_{ik} = \frac{1}{V} \int \sigma_{ik} dV.
\end{align}
Assuming the suspending fluid to be Newtonian, we may split the total volume integral into contributions from the fluid and from the particles,
\begin{align}
\bar{\sigma}_{ik} = \frac{1}{V} \int_{V - \sum V_o} \left \{ -\mathcal{P}\delta_{ik} + \mu \left(\frac{\partial U_i}{\partial x_k}+\frac{\partial U_k}{\partial x_i}\right) \right \}dV + \frac{1}{V} \sum \int_{V_o} \sigma_{ik} dV
\label{eqn:totstress}
\end{align}
where $V_o$ is the volume occupied by each active particle and the sum is over active particles. We next define a volume averaged velocity gradient 
\begin{align}
\frac{\partial u_i}{\partial x_k} = \frac{1}{V} \int_V \frac{\partial U_i}{\partial x_k} = \frac{1}{V} \int_{V-\sum V_o} \frac{\partial U_i}{\partial x_k} + \frac{1}{V} \sum \int_{A_o} U_i n_k dA
\label{eq:velreplace}
\end{align}
that will contribute to the deviatoric part of the stress tensor, in the absence of active particles, but with same velocity gradient $\nabla \mathbf{U}$ in the fluid regions. In the last step in  (\ref{eq:velreplace}) the divergence theorem has been used to replace the volume integral over each particle by a surface integral.

We also have the following mathematical relationship: 
\begin{align}
\int_{A_o} \sigma_{ij} x_k n_j dA  = \int_{V_o} \frac{\partial}{\partial x_j} (\sigma_{ij} x_k) dV = \int_{V_o} \frac{\partial \sigma_{ij}} {\partial x_j} \;x_k \; dV + \int_{V_o} \sigma_{ik}  dV = \int_{V_o} \sigma_{ik}  dV 
\label{eq:surfstress}
\end{align}
where we use $\partial \sigma_{ij}/\partial x_j = 0 $, assuming that the active particles are rigid and not accelerating.  This allows us to rewrite the internal stress of each particle as a surface contribution.

Using eqs.~(\ref{eq:velreplace}) and  (\ref{eq:surfstress}) the mean stress tensor in eq.~(\ref{eqn:totstress}) may be written 
\begin{align}
\bar{\sigma}_{ik} = -\frac{\delta_{ik}}{V} \int_{V - \sum V_o} \mathcal{P}dV &+ \mu \left(\frac{\partial u_i}{\partial x_k}+\frac{\partial u_k}{\partial x_i}\right) \nonumber\\
&+ \frac{1}{V} \sum \int_{A_o} [- \mu(U_i n_k + U_k n_i) + \sigma_{ij} x_k n_j ] dA.
\label{eqn:totstresssimp}
\end{align}
The first of the terms in eq.~(\ref{eqn:totstresssimp}) is a purely isotropic contribution to the pressure. The second term is proportional to the symmetric part of the velocity gradient tensor, and hence is the Newtonian contribution to the stress tensor.  
For rigid active particles, the translational and rotational velocities are constants at any given time and therefore the third term integrates to zero.
 The final term, that corresponds to the first moment of the stress distribution, will give rise to the active stress introduced in eq.~(\ref{eqn:actistress}).
The symmetric part of this first moment is known as a stresslet in the context of the multipole expansion of the Stokes equation for a force distribution acting on a fluid \cite{KimandKarrila}.
%
%

We now calculate the stress contributed by the final term in eq.~(\ref{eqn:totstresssimp}) for a suspension of active particles.
By definition, active particles generate motion autonomously, with no external forces acting. Therefore the simplest way to model the force distribution of an active entity is as a force dipole. We consider two co-linear, equal and opposite forces 
$\pm f \mathbf{p}$ that act at a distance $d \mathbf{p}$ apart, along an orientation vector $\mathbf{p}$ that characterises the active particle. Then
%
%
%
\begin{align}
\int_{A_o} \sigma_{ij} x_k n_j dA = fd(p_i p_k)
\end{align}
and eq.~(\ref{eqn:totstresssimp}) may be written 
\begin{align}
\bar{\sigma}_{ik} = -P \delta_{ik} + \mu \left (\frac{\partial u_i}{\partial x_k}+\frac{\partial u_k}{\partial x_i}\right) + \frac{1}{V} \sum fd(p_i p_k - \frac{1}{3} \delta_{ik})
\end{align}
where the contribution $\frac{1}{V} \sum fd \frac{1}{3} \delta_{ik}$ has been accommodated in the modified pressure $P$. Since, by definition, 

\begin{equation}
 \frac{1}{V} \sum (p_i p_k - \frac{1}{3} \delta_{ik}) \propto Q_{ik}
\end{equation}
we may write the constitutive relation for an active suspension as
\begin{align}
\bar{\sigma}_{ik} = - P \delta_{ik} + \mu \left(\frac{\partial u_i}{\partial x_k}+\frac{\partial u_k}{\partial x_i}\right) - \zeta Q_{ik}.
\end{align}
The activity coefficient, $\zeta$, determines the strength of the active stress. It can take either sign, depending on the details of the force distribution exerted on the fluid by the active particles. Systems with positive $\zeta$ are referred to as extensile or pushers whereas those with negative  $\zeta$ are referred to as contractile or pullers.  Extensile rod-shaped bacteria, such as Escherichia coli, pump fluid outwards from their ends and inwards towards their sides, whereas the contractile alga Chlamydomonas pumps fluid inwards towards its ends and outwards from its sides.
 Mixtures of molecular motors and cytoskeletal elements also produce different types of stress. For example, myosin, dynein or kinesin generate contractile stresses as they move on actin filaments or microtubules. However, depletant molecules may cause the formation of microtubule bundles in a suspension and two-headed kinesin molecular motors cause the filaments to slide past each other generating extensile stresses \cite{Dogic2012, Shelley2014}. Cell division is another source of extensile stress \cite{Amin2015cells}.


In the next section we show that the active stress leads to hydrodynamic instabilities that destabilise nematic ordering.

\subsection{Stability analysis}

\begin{figure}
\center
\resizebox{0.7\columnwidth}{!}{\includegraphics{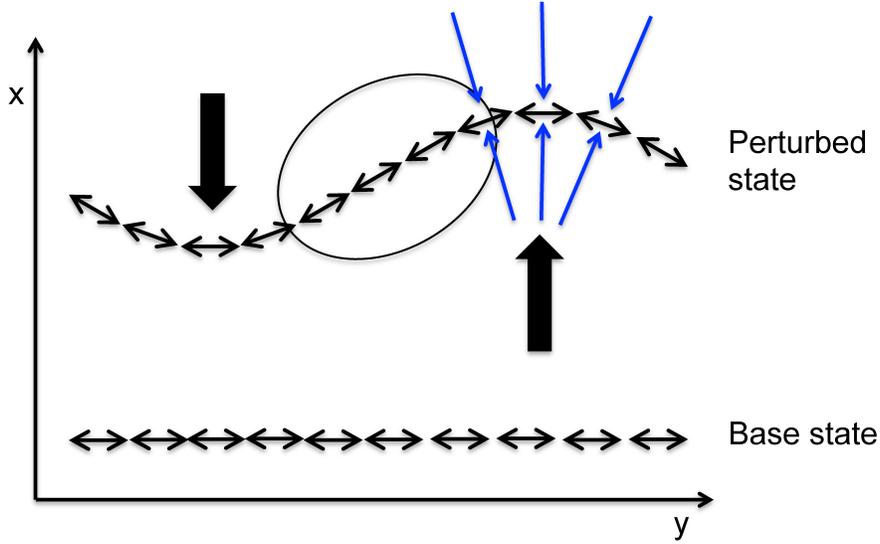}}
\caption{Schematic illustration of the hydrodynamic instability in active nematics \cite{Madan2007}. Extensile active particles are initially oriented in the $y$ direction. A bend perturbation in the director field with wavevector along $y$ generates flow along $x$. The resultant velocity gradients act to magnify the perturbation: the director field in the encircled region responds to the shear by rotating counter clockwise.}
\label{fig:instability}
\end{figure}

To demonstrate the inherent hydrodynamic instability of the active nematohydrodynamic equations we follow the calculations in \cite{Madan2007}, but to simplify the analysis we assume that the magnitude of the order parameter is constant everywhere. This substantially reduces the complexity of eqns.~(\ref{eqn:lc}) and (\ref{eqn:ns}) to 
\begin{align}
\partial_t n_i + u_j \partial_j n_i &= \lambda_1 E_{ij} n_j + \Omega_{ij} n_j + \Gamma_1 h_i ,\label{eqn:lcle}\\
\rho (\partial_t u_i + u_j \partial_j u_i) &= \partial_j \Pi_{ij} ;~~~~\partial_i u_i = 0,\label{eqn:nsle}
\end{align}
where $\mathbf{h} = K \nabla^2 \mathbf{n}$ is the molecular field. The passive and active parts of the stress tensor, (\ref{eqn:passstress}) and (\ref{eqn:actistress}), become
\begin{align}
\Pi_{ij}^{passive} &= -\frac{\lambda_1}{2} (n_i h_j + n_j h_i) + \frac{1}{2} (n_i h_j - n_j h_i), \\
\Pi_{ij}^{active} &= \zeta_1 (n_i n_j - \frac{\delta_{ij}}{3}),
\end{align}
where $\lambda_1 = \frac{3q+4}{9q}\lambda$, $\Gamma_1 = \frac{2}{9q^2}\Gamma$ and $\zeta_1 = \frac{3q}{2}\zeta$ are modified coefficients \cite{Davide2007}. 

We assume that, in the unperturbed state, the active constituents are ordered along the $y$-direction and that the velocity field is zero everywhere. We consider a long wavelength bend perturbation to this aligned structure, with wavevector along $\hat{y}$, that tilts the active particles slightly into the $x$-direction. The consequent imbalances in the active flows will generate a velocity field along $x$,
 $\mathbf{u} = u_x(y) \mathbf{\hat{x}} $, as illustrated in  Fig.~\ref{fig:instability}.

Considering only variations in the $y$-direction, eqs.~(\ref{eqn:lcle}) and (\ref{eqn:nsle}) simplify to
%
\begin{align}
\partial_t n_x &= \frac{1}{2} (\lambda_1 -1) \frac{d u_x}{dy} n_y + \Gamma_1 h_x,
\label{eqn:instab1}\\
\mu \frac{d^2 u_x}{dy^2} &= \zeta_1 \partial_y (n_x n_y) - \partial_y \left ( - \frac{\lambda_1}{2} (n_x h_y + n_y h_x) + \frac{1}{2} (n_x h_y - n_y h_x) \right).
\label{eqn:instab2}
\end{align}
Neglecting  terms in the molecular field, as higher order in derivatives, eq.~(\ref{eqn:instab2}) can immediately be integrated and substituted into
eq.~(\ref{eqn:instab1}) to give
\begin{align}
\partial_t n_x = \frac{\zeta_1}{2\mu} (\lambda_1-1) n_x,
\label{eqn:stab}
\end{align}
noting that, to linear order, $n_y=\sqrt{1-n_x^2}\approx 1$.
For rod-like ($\lambda_1 >1$), extensile ($\zeta_1 > 0$) particles, eq.~(\ref{eqn:stab}) predicts an exponential growth for bend perturbations.  If instead we assume a splay perturbation in the $y$-direction, which generates a flow along $\mathbf{\hat{y}} $, the nematic ordering of rod-like, contractile particles is unstable. 

A fluctuation will, in general, contain both splay and bend components and hence this analysis shows that nematic ordering in a wet active system is inherently unstable. This was first recognised by \cite{Sriram2002}. Very similar calculations leading to the onset of activity driven flow in a channel are presented in \cite{Joanny2005} . 

\subsection{Numerical approaches}

The analysis in Sec.~2.2 showed that the nematic state of active particles is hydrodynamically unstable. However, the complexity of the equations limits the feasibility of analytical results, and therefore numerical solutions of the equations of motion are needed to help understand the dynamical state of active turbulence that is a consequence of the instabilities. There have been several studies where the governing equations were solved numerically in various contexts and geometries  \cite{Davide2007, Miha2013, Cates2009}. The rheology of active materials has also been addressed \cite{Davide2007B, Suzanne2011} as has the role of topological defects in active systems  \cite{Mahadevan2011, ourprl2013, Giomi2013}.

Different methods may be used to solve the active nematohydrodynamic equations. A hybrid lattice Boltzmann approach has proved particularly useful, and has been applied in several studies by us and other groups \cite{Davide2007, Miha2013, Cates2009, Suzanne2011}. In this scheme eq.~(\ref{eqn:ns}) is solved using a standard lattice Boltzmann method.  Eq.~(\ref{eqn:lc}) is spatially discretised using a central difference scheme, and the resulting set of ordinary differential equations are integrated in time using an Euler approach. The solutions are coupled at each time step by updating the order parameter field to calculate the active and passive parts of the stress field and by updating the velocity field in order to calculate the convective and co-rotational derivatives. More details about this hybrid solution method can be found in \cite{ourpta2014, Davide2007}. An alternative way to solve the Navier Stokes equations is the stream function--vorticity formulation described in \cite{Giomi2014}. 

To date, most experiments and numerical studies have considered two dimensions. We describe results from the simulations below.

\section{The onset of active turbulence}

\begin{figure}
\resizebox{\columnwidth}{!}{ \includegraphics{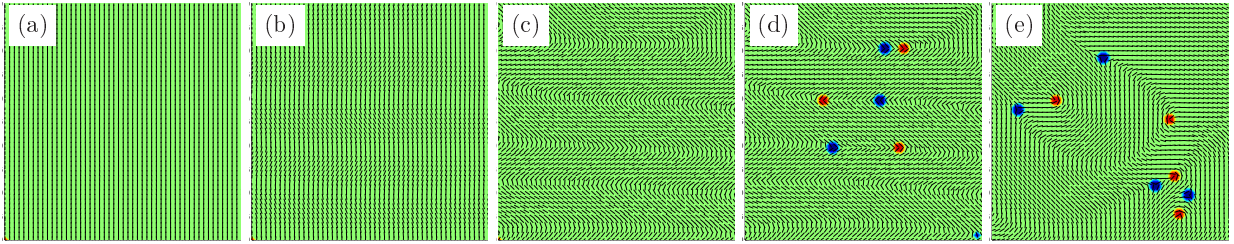}} 
\caption{Development of active turbulence: (a) Nematic order of the director field (b) Onset of the hydrodynamic instability (c) Formation of parallel walls, lines of bend deformation separating nematic regions  (d) Decay of the walls to pairs of topological defects that move apart along the walls (e) Wall bending, and the onset of fully developed active turbulence. (After \cite{ourepl2014}) }
\label{fig:start}       
\end{figure}

Fig.~\ref{fig:start}  describes how the hydrodynamic instability of the nematic state in an active material develops into active turbulence. The panels in the figure show snapshots of the evolution in time of the director field, calculated from numerical integration of the active nematic equations of motion (\ref{eqn:lc}) and (\ref{eqn:ns})  \cite{ourepl2014}. In the hybrid lattice Boltzmann approach used in \cite{ourepl2014}, the following parameters were used, $\Gamma = 0.34, A = 0.0, B = - 0.3, C = 0.3, K = 0.02, \mu = 2/3, \lambda=0.7$ and $\zeta = 0.0125$ (extensile) . These parameters are reported in lattice units where discrete time and space steps are chosen as unity. The simulation is started with the director field in a nematic state with slight perturbations as shown in panel (a).  The perturbations grow due to the hydrodynamic instability  (panel b). These wave-like deformations in the director field localise to form {\bf walls} as shown in panel (c). Walls are lines of bend deformations which separate nematic regions. In panel (c) they are parallel and equidistant, but this is a transient state. The walls start to undulate, and pairs of oppositely charged, $\pm {1}/{2}$, topological defects are formed at the walls by a combination of elastic and active stresses (panel d). Gradients in the nematic field around a defect generate active stresses and hence flow which separates the defect pair. The defects initially move along the wall but their dynamics, together with that of the walls, rapidly becomes chaotic. Hence defects of opposite charge meet and annihilate leading to a steady state, the fully developed active turbulence shown in panel (e). 


Fully developed active turbulence is described schematically in Fig. \ref{flowfig}(e). Hydrodynamic instabilities give rise to walls that fluctuate and decay into topological defects. Defects move apart, and annihilate with other defects, in a way that tends to restore nematic order. This cycle of  the formation and destruction of nematic regions (or equivalently of walls and defects), driven by the activity, sustains the state of active turbulence. 


\begin{figure}
\center
\subfigure[]{\includegraphics[trim = 0 0 0 0, clip, width=0.48\linewidth]{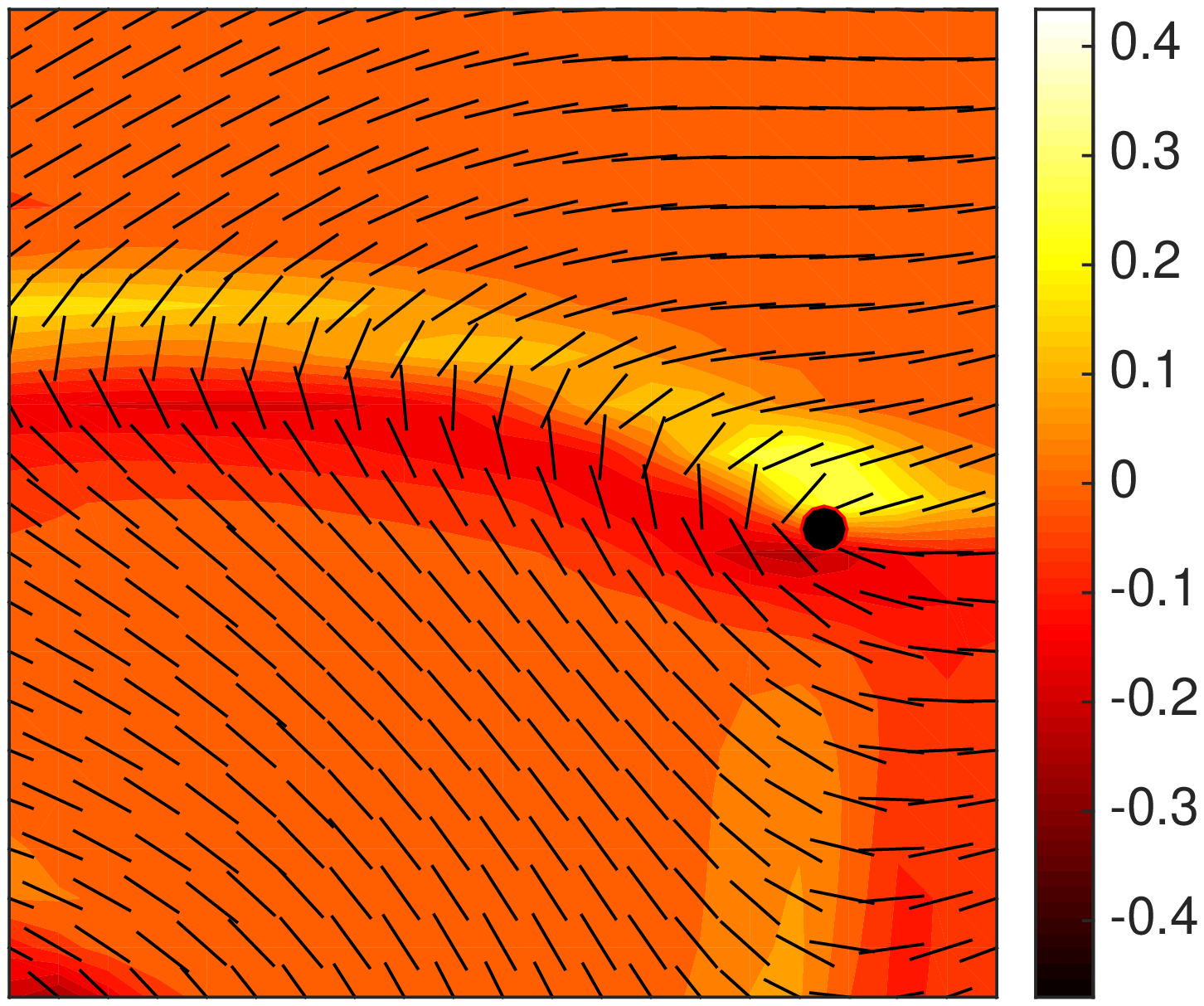}}
\subfigure[]{\includegraphics[trim = 0 0 0 0, clip, width=0.5\linewidth]{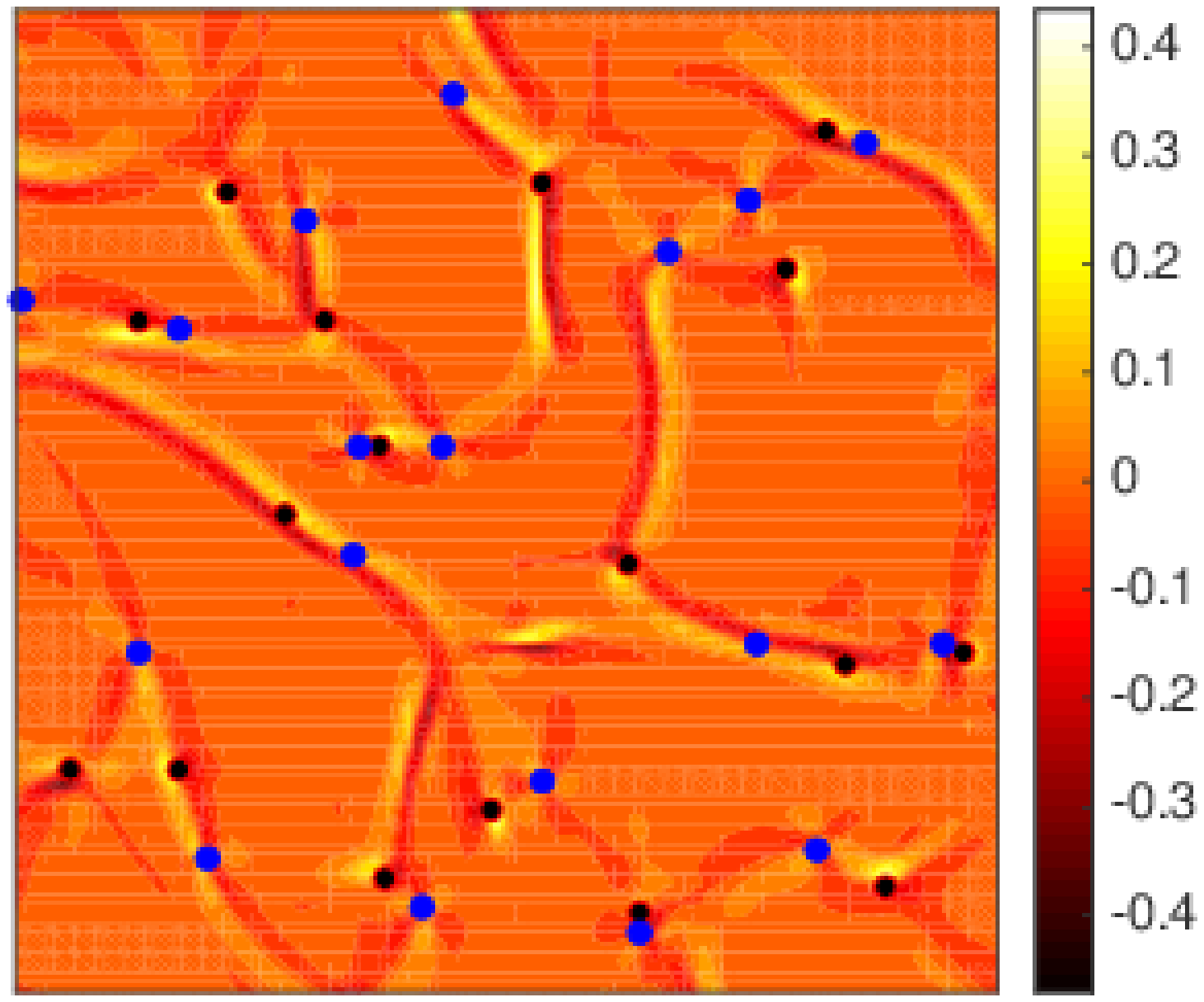}}
\caption{(a) Illustration of a wall and a topological defect. The background colour illustrates the value of $\nabla \wedge \nabla \cdot \mathbf{Q} $ from yellow (high) to red (low).
(b) A larger domain showing that the quantity $\nabla \wedge \nabla \cdot \mathbf{Q} $  highlights the walls and topological defects, shown in black ($+1/2$) and blue ($-1/2$).}
\label{fig:curlq}       
\end{figure}


Fig.~\ref{flowfig} (b-d) also illustrates the velocity and vorticity fields associated with the defects and walls in the director field. To relate the vorticity to gradients in $\mathbf{Q}$, we neglect elastic stresses in the system (which are usually small), and  balance the viscous and active forces to give
\begin{align}
\mu \nabla^2 \mathbf{u} \sim \zeta \nabla \cdot \mathbf{Q}.
\end{align}
Taking the curl of both sides, and solving for the 2D vorticity $\omega$, 
\begin{align}
\mathbf{\omega} (\mathbf{r}) = \frac{\zeta}{2 \pi \mu} \int_{S} \nabla \wedge \nabla \cdot \mathbf{Q} (\mathbf{r}') \ln|\mathbf{r}-\mathbf{r'}| dS.
\label{eqn:vort}
\end{align}
Eq.~(\ref{eqn:vort}) suggests that regions with $ \nabla \wedge \nabla \cdot \mathbf{Q}\neq 0$ are the locations at which vorticity is generated. Fig.~\ref{fig:curlq} shows that this quantity is large at the walls and topological defects characterising deformations in the director field. Thus walls and defects can be identified as sources of vorticity. 
The vorticity then diffuses within the domain to generate active turbulence. By contrast, in high Reynolds number turbulence, vorticity is advected in the fluid through mechanisms associated with the inertial terms in the Navier Stokes equation 
\cite{Bratanov2015}.


%
%
%
%
%
\section{Discussion}

We have described the underlying mechanisms that sustain active turbulence, but a predictive theory is still lacking. 
Scaling arguments relating the velocity, vorticity and number of topological defects are given in \cite{ourpta2014}. Giomi \cite{Giomi2015} hypothesised an exponential distribution of vortices to calculate both the spectral behaviour of the flow field and the mechanisms behind the dynamics of topological defects. However, whether the flow, vorticity or director field evolution should be considered the primary driver of active turbulence, and whether walls or defects are the primary excitations,  remain as open questions. Some recent studies have focussed on reducing the complexity of the equations to give minimal models \cite{Jorn2013, Putzig2015, Oza2015}. Others \cite{Shelley2014, Gao2015} aim to incorporate the microscopic mechanisms that yield an active stress, thus directly taking the polar nature of the active constituents into consideration. In all these approaches the qualitative features of active turbulence remain unchanged.

At low activities, fluctuating walls are still formed. However, there is insufficient energy for them to decay into topological defects. Nevertheless, since walls can generate flow fields, active turbulence is, in general, still observed. In smaller systems, however, time-independent flow patterns can be stabilised \cite{Joanny2005, Goldstein2012, Miha2013, Wioland2013, Lushi2014}. For slightly larger system sizes (or, equivalently, activities) the system can attain an oscillatory behaviour, with alternating directions of the walls
 \cite{Giomi2011}. 

It is of interest to ask which features of active turbulence are universal: there are obvious visual similarities in experiments across a wide range of length scales, from suspensions of filaments and molecular motors and bacteria to flocks of birds and schools of fish. In the dry limit the mechanism of hydrodynamic instability that leads to active turbulence is lost. However topological defects have been observed in experiments \cite{Narayan2007} and simulations \cite{Shi2013} on dry systems, possibly as a consequence of details of the interactions between active particles.  
Recent simulations \cite{Amin2015} show that substrate friction, which screens the long ranged hydrodynamics \cite{Sumesh2014}, can be tuned to connect the two limits of wet and dry systems. In the crossover regime, lattices of flow vortices and topological defects are stabilised as the hydrodynamic length scale is reduced to that of the vortex created by a defect. 
The stabilisation of vortices and order in the microstructure opens a gateway to the design of micromachines to harness energy from active systems.\\

We acknowledge funding from the ERC Advanced Grant 291234 MiCE.  We thank  Andrew Balin, Matthew L Blow, Amin Doostmohammadi,  Jorn Dunkel,  Ramin Golestanian, Mitya Pushkin, and Tyler Shendruk for helpful discussions.

%
%
%
%
%
%
\bibliography{refe.bib}{}

%

\end{document}